\documentstyle[12pt,epsfig,fleqn]{article} \oddsidemargin .4cm 
\newcommand{\be}{\begin{eqnarray}} 
\newcommand{\ee}{\end{eqnarray}} 
 \newcommand{\nn}{\nonumber}

\begin{document} 
\begin{titlepage} 
\begin{flushright} INRNE-TH-97/08\\ 
September 1997 \end{flushright} \vfill 
\begin{center} 
{\Large\bf On the top-quark polarization  and \\ 
how to measure it}\\ 
\vspace{1.5cm} {\large Ekaterina Christova} \\
\vspace{0.7cm} 
{\em Institute of Nuclear Research and  Nuclear Energy}\\ 
{\em Boul. Tzarigradsko Chaussee 72, Sofia 1784, Bulgaria}\\ 
\vspace{1cm}and\\ \vspace{1cm} 
{\large Dimitar Draganov} \\ 
\vspace{0.7cm} {\em Physics Department, Sofia University}\\ 
{\em  Boul. James Baucher  5, Sofia 1126, Bulgaria} \vspace{1.5cm}  
{\sc Abstract} \end{center} 
The top quark is expected to decay as a free particle with 
definite momentum and polarization. Here we consider the 
possibilities to obtain information about its polarization 
through the energy and angular distributions of the b-quarks from  
the decay of the top-quarks produced in $e^+e^-$ annihilation. 
Analytic 
expressions and numerical estimates for different beam 
polarizations are derived. 
 
\end{titlepage} \newpage \setcounter{footnote}{0} \setcounter{page}{1} 
 
{\bf 1.} The large mass  of the top quark measured at  
Fermilab~\cite{CDF} allows to probe  physics at 
high energies where new physics is expected to appear as well. 
 Many papers treat this idea. That is why it becomes of basic 
interest  to test first of all the Standard Model (S.M.) for 
processes with top quarks.

There is one basic difference of the top-quark as compared to the 
other quarks: due to its  high 
mass, the top quark should decay before forming a hadronic bound 
state~\cite{Bigi}.  This means that 
the top quark will decay with a definite 
 polarization, that can be measured through the  
 distribution of its decay products, and that the  
 theoretical predictions  
become more reliable free of the complications of hadronization. 
 Measuring the top-quark polarization and comparing it 
to the  theoretical predictions would present a clear test 
of the S.M.  
 
All $t\bar t$ events contain a $b$ and a $\bar b$ -quark. 
Different methods for detecting the $t\bar t$ events, based on 
tagging b-jets (for example via high $p_{\perp}$ leptons from 
$b\to c(i^{\pm}\nu )$ or via displaced vertices of $b$ and $\bar 
b$-decays) exist~\cite{exp}.  
 
Here we discuss the possibilities to answer the questions whether 
the $t$-quark decays keeping its polarization and whether this  
polarization is the one predicted by S.M.,   
measuring the energy and angular distributions of the $b$-quarks 
  in the inclusive process: 
 \be 
 e^+(q_l) + e^-(q_{\bar l}) \to t(p_t) + \bar t (p_{\bar t}) 
\to b(p_b) + X\,.\label{1} 
\ee 
 $X$ stands for $W\bar t$. 
We assume that the initial beams are longitudinally polarized. 
 
As the $t$-quark actually does not mix with other quarks, the 
considered decay mode $t\rightarrow bW$  
 is  the only decay mode in S.M. This 
implies that the rate for production of $b$-quarks in (\ref{1}) 
will be  the same as the rate for production of $t$-quarks in 
${e^+e^-}$ annihilation. 
 
Previously the $t$-quark polarization was considered in 
refs. \cite{Dalitz}, \cite{Kane} and \cite{Sehgal}. 
It was shown that the S.M. predicts a rather large  
 polarization in the production plane of 
the $t$-quarks produced in the annihilation process of  
unpolarized $e^+e^-$.  The possibilities to measure  
it through the angular distribution of the leptons from  
the  decay mode $t\to bW \to bl\nu$ 
were considered in  \cite{Dalitz}, \cite{Sehgal} and \cite{Kuhn}.  
 
Here we obtain analytic 
expressions for the energy and angular distributions of the 
$b$-quarks in the c.m.system of the sequential process (\ref{1}). 
We follow the formalizm of \cite{BG}, in which the $t$-quark 
polarization enters explicitely.  Different energy and angular 
 asymmetries are defined.  
We show that measurements of the  energy and angular distributions of 
the $b$-quarks  would allow to determine  
 the $t$-quark polarization in the production plane. 
 Numerical estimates  for different beam 
polarizations  and the  
dependence on the incertainties in $m_t$ are presented. 
 %%%%%%%%%%%%%%%%%%%%%%%%%%%%%%%%%%%%%%%%%%%%%%%%%%%%%%%%%%%%%%%%%% 
 
 %%%%%%%%%%%%%%%%%%%%%%%%%%%%%%%%%%%%%%%%%%%%%%%%%%%%%%%%%%%%%%%%% 
\vspace{1.00cm} 
{\bf 2.} Following  the formalizm developed 
in~\cite{BG}, for the cross section $d\sigma_{\lambda\lambda'}$ 
of the sequential process (\ref{1}), $\lambda$ and $\lambda'$ being 
the longitudinal polarizations of $e^-$ and $e^+$,  we obtain: 
\be 
\frac{d\sigma_{\lambda\lambda'}}{d\cos\theta d\Omega_b} 
=\left(\frac{d\sigma_{\lambda\lambda'}}{d\cos\theta d\Omega_b}\right)_0 
\left(1 + \alpha_b m_t\frac{(\xi p_b)}{(p_t 
p_b)}\right)\,.\label{sigmacms}  
\ee 
 All quantities in 
(\ref{sigmacms})  are in the c.m.system, $\xi$ is the  
polarization 4-vector of the $t$-quarks   
determined in the production process $e^+e^-\rightarrow t\bar 
t$, $\cos\theta$ is the production  
angle of the $t$-quarks (Z 
points the direction of $e^-$), $\Gamma$ 
is the total decay width of the $t$-quark,  
\be  
\alpha_b =\frac{m_t^2-2m_W^2}{m_t^2+2m_W^2} 
\ee 
and its value determines the sensitivity of the b-quarks to the 
polarization of the $t$-quarks, $ (p_tp_b) = 
(m_t^2-m_W^2)/2$,  
\be 
\left(\frac{d\sigma_{\lambda\lambda'}}{d\cos\theta d\Omega_b}\right)_0 
  = \frac{\alpha^3_{\it 
em}}{32\sin^2\theta_w}\vert U_{tb}\vert^2\:\frac{3\beta}{s} 
\:\frac{2m_W^2 + m_t^2}{m_W^2} 
\:\frac{E^2_b}{m_t\Gamma}\:N_{\lambda\lambda'}\label{sigma0} 
\ee 
is the S.M. cross section for totally 
depolarized $t$-quarks, $\sqrt s$ is the 
c. m. energy, $\beta = \sqrt{1-4m_t^2/s}$, $U_{tb}$ is the 
corresponding element of the Kobayashi-Maskawa matrix,    
 $N_{\lambda\lambda'}$ equals 
\be 
N_{\lambda\lambda'}= \left( 1+\beta^2\cos^2\theta \right) F_1 
+\frac{4m_t^2}{s}F_2+2\beta\cos\theta F_3\,,\label{N} \ee 
where 
\be 
F_i = (1-\lambda\lambda')F_i^0 + (\lambda - \lambda')G_i^0\,. 
\ee 
Here 
\be 
&&F_{1,2}^0 = \left(\frac{2}{3}\right)^2 + h_Z^2(c_V^2+c_A^2)
(g_V^2 \pm g_A^2)-\frac{4}{3}h_Zc_Vg_V\nn\\ &&G_{1,2}^0 
=2h_Z^2\,c_Vc_A\left(g_V^2 \pm g_A^2\right)-\frac{4}{3} 
h_Z\,c_Ag_V\,,\label{FG}\\ 
&&F_3^0 = 4 h_Z c_Ag_A\left(h_Zc_Vg_V-\frac{1}{3}\right),  
G_3^0 =2h_Z^2\,\left(c_V^2+c_A^2\right)g_Vg_A-\frac{4}{3} 
h_Z\,c_Vg_A\,,\nn 
\ee 
$E_b$ is the energy of the b-quarks in the c.m.system: $E_b = 
(p_tp_b)/[E(1-\beta\cos\theta_{tb})]$,  
E is the energy of the initial leptons, $\cos\theta_{tb}$ is the 
decay angle  
 between the directions of the t and  b-quarks, 
   $c_V=-1/2+2\sin^2\theta_W\,,c_A=1/2$ and  
$g_V=1/2-4/3\,\sin^2\theta_W\,, 
g_A=-1/2$ are 
 the S.M. couplings of $Z^0$ to the 
electron and  the top-quark respectively, 
$h_Z=s/[\sin^22\theta_w(s-m_Z^2)]$.

%%%%%%%%%%%%%%%%%%%%%%%%%%%%%%%%%%%%%%%%%%%%%%%%%%%%%%%%%%%%%%%%%%%%% 
%%%%%%%%%%%%%%%%%%%%%%%%%%%%%%%%%%%%%%%%%%%%%%%%%%%%%%%%%%%%%%%%%%%% 
\vspace{1.00cm} 
{\bf 3.} Now we obtain the expression for the polarization 4-vector 
$\xi^\alpha$ of the top-quarks that are produced in  
 $e^+e^-$ annihilation with longitudinally polarized initial  
leptons in  tree-level approximation of the S.M.

As by definition  $(\xi p_t) =0$,  
 $\xi_\alpha$ in general (in tree level)  
can be decomposed along two independent 4-vectors, orthogonal to $p_t$, 
 that lay in the production plane. We choose them to be $Q_l$ and 
 $Q_{\bar l}$: 
\be   Q_l^{\alpha} = q_l^{\alpha} - \frac {p_tq_l}{m_t^2}\cdot 
p_t^{\alpha} 
\,,\qquad Q_{\bar{l}}^{\alpha} = q_{\bar{l}}^{\alpha} 
- \frac{p_tq_{\bar{l}}}{m_t^2}\cdot p_t^{\alpha}\,. \ee  
Then we write: 
\be (\xi)^\alpha 
={P}_l(Q_l)^\alpha + {P}_{\bar{l}}(Q_{\bar{l}})^\alpha 
\,.\label{xi1} \ee 
 
Using the method  of ~\cite{BG} we obtain: 
\be 
P_{l,\bar l}(\theta) =\pm \frac{2m_t}{s}\{(1\mp\beta\cos\theta ) 
(G_1\mp G_3) + 
(1\pm \beta\cos\theta )G_2\}/N_{\lambda\lambda'} 
\ee 
Here  
\be 
G_i = (1-\lambda\lambda')G_i^0 + (\lambda - \lambda')F_i^0\,, 
\ee 
$F_i^0$ and $G_i^0$ are given by (\ref{FG}). 
 
Having the explicit expression for $\xi^\alpha$, 
from (\ref{sigmacms}) one readily  obtains 
the energy and angular distributions of the $b$-quarks .  
 
There are two independent components of the polarization vector 
of the top in tree level of the S.M. Thus we need at least two  
independent measurements in order to get information about 
 $\xi^\alpha$. 
As it is evident from (\ref{sigmacms}) the polarization of 
 the $t$-quarks 
 1) introduces a new dependence on 
$\cos\theta_b$, not only though $\cos\theta_{tb}$,  and 2)    
changes the behaviour  
 on $\cos\theta_{tb}$, or equivalently on $E_b$. We shall 
explore both to get information about $\xi$.  
 
%%%%%%%%%%%%%%%%%%%%%%%%%%%%%%%%%%%%%%%%%%%%%%%%%%%%%%%%%%%%%%%%%%%% 
%%%%%%%%%%%%%%%%%%%%%%%%%%%%%%%%%%%%%%%%%%%%%%%%%%%%%%%%%%%%%%%%%%%%%% 
\vspace{1.00cm} 
{\bf 4.} For the energy spectrum of 
 the  $b$-quarks in the c.m.s. we obtain: 
\be 
\frac{d\sigma }{dE_b}=\left(\frac{d\sigma 
}{dE_b}\right)_0\left [1+ 
\frac{4\alpha_b G_3}{(3+\beta^2)F_1 + 3(1-\beta^2)F_2} 
\left(1-\frac{4m_t^2}{m_t^2-m_W^2} 
\,\frac{E_b}{\sqrt s}\right)\right ]\label{Eb} 
\ee 
where 
\be 
\left(\frac{d\sigma}{dE_b}\right)_0 &=& 
\frac{\alpha^3_{em}}{8\sin^2\theta_w\, s\sqrt s}\, 
\pi\vert U_{tb}\vert^2\: 
\frac{m_t^2 + 2m_W^2}{m_W^2}\,\frac{m_t^2 -m^2_W}{m_t 
\Gamma}\times\nn\\ 
&&\times [(3 + \beta^2)F_1 + 3(1 -\beta^2)F_2] 
\ee 
is the energy spectum for unpolarized $t$-quarks. 
Thus, if the $t$-quarks decay depolarized, the cross section for a 
given energy $E_b$ would  
 not depend  on the energy of the $b$-quark. The polarization of 
the decaying $t$-quarks turns this constant behaviour into a 
linear $E_b$-behaviour, the decline determined by $G_3$. 
 This allows to form different asymmetries, whose nonsero value 
would imply a nonzero polarization of the $t$-quarks. 
 
If $N_{\lambda\lambda'}(E_b)$ is the number of $b$-quarks with 
energy $E_b$, then the ratio $R(\Delta E_b)$:   
\be 
R_{\lambda\lambda'}(\Delta E_b)= 
\frac{N_{\lambda\lambda'}(E_2)-N_{\lambda\lambda'}(E_1)} 
{N_{\lambda\lambda'}(E_2)+N_{\lambda\lambda'}(E_1)}\,, 
\qquad \Delta E_b = E_2 - E_1\label{deltaR} 
\ee 
would be  
proportional to the top-polarization. 
As the energy spectrum exibits a linear dependence on the 
polarization, $R(\Delta E_b)$ would reach its maximum if 
measurements are fulfilled at the ends of the spectrum: 
 $E_2= E_{max}\,, E_1=E_{min}$. Then we obtain: 
\be 
R_{\lambda\lambda'}(E_{max}-E_{min})=\frac{-4\alpha_b\beta G_3} 
{(3 + \beta^2)F_1 + 3(1 -\beta^2)F_2}\,.\label{deltaRmax} 
\ee 
 
One can form different integral asymmetries comparing the number 
of $b$-quarks in two energy intervals with the same length. We 
consider the following asymmetry, whose  nonzero value would  
an indication of polarized decaying $t$-quarks.: 
\be 
{\cal R}_{\lambda\lambda'}=\frac{N_{\lambda\lambda'} 
(E_b\,>\,{\cal E}_0)- 
N_{\lambda\lambda'}(E_b\,<\,{\cal E}_0)} 
{N_{\lambda\lambda'}(E_b\,>\,{\cal E}_0)+N_{\lambda\lambda'} 
(E_b\,<\,{\cal E}_0)}= 
\frac{-2\alpha_b\beta G_3}{(3+\beta^2)F_1 + 3(1-\beta^2)F_2}\,. 
\label{calRmax} 
\ee 
where   ${\cal E}_0$ is the mean energy:  
${\cal E}_0=(E_{min}+E_{max})/2 = 
\sqrt s (m_t^2-m_W^2)/(4m_t^2)$,  
$N_{\lambda\lambda'}(E_b\,>\,{\cal E}_0)$ and  
$N_{\lambda\lambda'}(E_b\,<\,{\cal E}_0)$ 
are the number of $b$-quarks with energy  
$E_b>{\cal E}_0$ and  $E_b<{\cal E}_0$ 
respectively.  
 
%%%%%%%%%%%%%%%%%%%%%%%%%%%%%%%%%%%%%%%%%%%%%%%%%%%%%%%%%%%%%%%%%%%% 
%%%%%%%%%%%%%%%%%%%%%%%%%%%%%%%%%%%%%%%%%%%%%%%%%%%%%%%%%%%%%%%%%% 
 
\vspace{1.00cm} 
{\bf 5.} Integrating (\ref{sigmacms}) over $\theta$ and $\varphi_b$ 
 we obtain 
the  $\cos\theta_b$-distribution of the b-quarks in the 
c.m.system.  
We write it in a form 
that keeps in evidence the dependence on the beam polarization too: 
\be 
\frac{d\sigma_{\lambda\lambda'}}{d(\cos\theta_b)}  
&=&\frac{3\alpha^3_{em}}{64\, s\sin^2\theta_w}\:\pi 
\vert U_{tb}\vert^2\:\frac{(m_t^2 - m^2_W)^2(m_t^2 +2m^2_W)} 
{m_t^3 m^2_W\Gamma}\times\nn\\ 
&&\times \Big\{(1-\lambda\lambda') 
[C_0 +C_1\cos\theta_b +C_2\cos^2\theta_b]\nn\\ 
&&\hspace{0.5cm} 
+(\lambda-\lambda')[O_0 +O_1\cos\theta_b +O_2\cos^2\theta_b]\Big\} 
\label{costheta} 
\ee 
where 
\be 
C_0&=& F_1^0\left(2\beta^3 + (1-\beta^2)  
\ln\frac{1+\beta}{1-\beta}\right) + 
2\beta(1-\beta^2)F_2^0 \nn\\ 
&& -\alpha_b (1-\beta^2)G_3^0 
\left(2\beta - \ln\frac{1+\beta}{1-\beta} \right) 
\label{C0} 
\ee 
\be 
C_1& =& 2F_3^0\left(2\beta -(1-\beta^2)\:\ln\frac{1 +\beta} 
{1-\beta}\right)\nn\\ 
&& + 2\alpha_b(1-\beta^2)\left[\beta(G_1^0 - G_2^0) - 
G_1^0\:\ln\frac{1 +\beta} 
{1-\beta}\right]\label{C1} 
\ee 
\be 
C_2& =&F_1^0\left(2\beta(3 -2\beta^2) -3(1 -\beta^2)\: 
\ln\frac{1 +\beta}{1-\beta}\right)\nn\\ 
&& + 3\alpha_b (1-\beta^2)G_3^0\left(2\beta -\ln\frac{1 +\beta} 
{1-\beta}\right)\label{C2} 
\ee 
\be 
O_i=C_i\,(F_i^0 \longleftrightarrow G_i^0) 
\ee 
 
 As it is evident from (\ref{C0}) - (\ref{C2}),  
 the polarization of the $t$-quarks does 
not introduce any new type of $\cos\theta_b$-dependence. 
 Thus, the  angular-polarization  
asymmetries, contrary to the considered energy asymmetries,  
 would be nonzero for both polarized and depolarized 
$t$-quarks, but differ numerically.  
 
Let $\sigma_{\lambda\lambda'}^F$ and $\sigma_{\lambda\lambda'}^B$ 
denote the number of $b$-quarks produced in the forward and 
backward hemispheres respectively: 
\be 
\sigma_{\lambda\lambda'}^F=2\pi\int_{\theta_0}^{\pi /2} 
\left(\frac{d\sigma_{\lambda\lambda'}}{d\cos\theta_b}\right) 
\sin\theta_b\,d\theta_b\,, 
\quad \sigma_{\lambda\lambda'}^B=2\pi\int_{\pi /2}^{\pi -\theta_0} 
\left(\frac{d\sigma_{\lambda\lambda'}}{d\cos\theta_b}\right) 
\sin\theta_b\,d\theta_b\nn 
\ee 
$\theta_0$ is determined by the experimental set up. 
The forward-backward asymmetries that we shall consider are 
sensitive to the  combinations  
 $\left[\beta (G_1^0 - G_2^0) - 
G_1^0\:\ln\frac{(1 +\beta )}{(1-\beta )}\right]$ 
 and $\left[\beta (F_1^0 - F_2^0) - 
F_1^0\:\ln\frac{(1 +\beta )} 
{(1-\beta )}\right]$  that enter $C_1$ and  
$O_1$. The other contribution to $\xi^\alpha$, determined by  
 $G_3^0, F_3^0$ that 
enter $C_{0,2}, O_{0,2}$ can be measured by  the energy spectrum, 
as shown above.  
Thus, measuring the energy and angular distributions of the $b$- 
quarks, with either polarized or unpolarized initial beams 
  we obtain information about two independent contributions 
to the polarization of the $t$-quark. 
  
 We consider the following angular-polarization $b$-jet asymmetries: 
 
{\bf i)} The forward-backward asymmtry  
 $A^{FB}_{\lambda\lambda'}$, that measures the difference between 
the $b$-quarks in the forward and backward hemispheres: 
\be  
A^{FB}_{\lambda\lambda'} =\frac{\sigma_{\lambda\lambda'}^F 
-\sigma_{\lambda\lambda'}^B}{\sigma_{\lambda\lambda'}^F+ 
\sigma_{\lambda\lambda'}^B} 
\ee 
From (\ref{costheta}) for $A^{FB}_{\lambda\lambda'}$ we obtain: 
\be 
A^{FB}_{\lambda\lambda'} =\frac{(1-\lambda\lambda')\,C_1 + (\lambda 
-\lambda')\, O_1}{ (1-\lambda\lambda')\,(2\,C_0+2\,C_2/3) + 
(\lambda -\lambda')\, (2\,O_0 + 2\, O_2/3)}\label{AFB1}  
\ee 
 
Our expressions (\ref{costheta}) and (\ref{AFB1}) for the  
$\cos\theta_b$-distridution and for the forward-backward 
asymmetry (\ref{AFB1}) for unpolarized initial leptons 
($\lambda=\lambda' =0$)  
coincide with those obtained in~\cite{Sehgal}, where they were 
first derived using the technique of ref~\cite{Tsau}).

{\bf ii)} The polarization angular  
asymmetry ${\cal A}^{FB}$, that compares the  $b$-quarks in 
the forward and backward hemispheres with opposite beam polarizations: 
\be  
{\cal A}^{FB} =\frac{\sigma^F_{\lambda\lambda'}- 
\sigma^B_{-\lambda-\lambda'}} 
{\sigma^F_{\lambda\lambda'}+\sigma^B_{-\lambda-\lambda'}} 
=\frac{(1-\lambda\lambda')\,C_1 + (\lambda 
-\lambda')\,(2\,O_0+2\,O_2/3)}{(1-\lambda\lambda')\, 
(2\,C_0+2\,C_2/3) + (\lambda 
-\lambda')\,O_1}\label{calA1} 
\ee 
 
{\bf iii)} The polarization asymmetry ${\cal B}_{LR}^{FB}$ that 
singles out $O_1$ and $C_1$ only: 
\be 
{\cal B}^{FB}=\frac{1-\lambda\lambda'}{\lambda -\lambda'} 
\,\frac{\left(\sigma^F_{\lambda\lambda'}- 
\sigma^B_{\lambda\lambda'}\right) 
-\left(\sigma^F_{-\lambda-\lambda'}- 
\sigma^B_{-\lambda-\lambda'}\right)} 
{\left(\sigma^F_{\lambda\lambda'}-\sigma^B_{\lambda\lambda'}\right) 
+\left(\sigma^F_{-\lambda-\lambda'}-\sigma^B_{-\lambda-\lambda'} 
\right)}\,.=\frac{O_1}{C_1}\label{calB} 
\ee 
Note that ${\cal B}^{FB}$  is defined for polarized beams 
only and is independent  
on the beam polarization, ${\cal B}^{FB}(\lambda'=\lambda = 0) = 0$. 
 
The above asymmetries contain also the combinations $2\,C_0+2\,C_2/3$ and 
$2\,O_0+2\,O_2/3$. These combinations  enter the total cross section and  
are independent on the $t$-quark polarization. 
%%%%%%%%%%%%%%%%%%%%%%%%%%%%%%%%%%%%%%%%%%%%%%%%%%%%%%%%%%%%%%%%%%%%%%%%% 
%%%%%%%%%%%%%%%%%%%%%%%%%%%%%%%%%%%%%%%%%%%%%%%%%%%%%%%%%%%%%%%%%%%%%%%% 
 
%%%%%%%%%%%%%%%%%%%%%%%%%%%%%%%%%%%%%%%%%%%%%%%%%%%%%%%%%%%%%% 
\vspace{1.00cm} 
{\bf 6.}  
We have estimated the defined asymmetries  
%$R$, ${\cal R}$, $A^{FB}$, 
%${\cal A}_{LR}^{FB}$ and ${\cal B}_{LR}^{FB}$ 
  at  $\sqrt s=500$ GeV assuming $\lambda'=-\lambda$.  
  We examine their 
sensitivity to the uncertainties in $m_t$ for different beam 
polarizations. 
 
Observation of a linear $E_b$-dependence of the $b$-quark 
spectrum in stead of a constant one, or equivalently non-zero 
values of $R$ or ${\cal R}$ would be possible indications for  
polarized decaying $t$-quarks. 
Their values are determined by $G_3$.  
On Fig. 1 we show the dependence of $R(E_{max}-E_{min})$  
on the beam polarization $\lambda$. $R$ depends strongly on  
$\lambda$ - it varies from $\approx 6\%$ for $\lambda = 0$ up to 
$\approx 17\%$ for $\lambda = \pm 0.5$. As our analysis showed  
it is actually independent 
on $m_t$. The integral asymmetry ${\cal R}$ should be  two 
times less, as evident from (\ref{calRmax}).     
 
The sensitivity of the angular asymmetries $A^{FB}, {\cal A}^{FB}$ and 
${\cal B}^{FB}$ to $\xi$ is determined by the difference between their 
values $A^{FB}, {\cal A}^{FB}$ and 
${\cal B}^{FB}$ with polarized decaying $t$-quarks - eqs. 
(\ref{AFB1}), (\ref{calA1}) and (\ref{calB}), and the corresponding 
values ${\bar A}^{FB}, {\bar{\cal A}}^{FB}$ and 
$\bar{\cal B}^{FB}$ with depolarized decaying $t$-quarks ($\alpha_b=0$). 
It is determined by $G_1$ and $G_2$. 
 
On Fig. 2, the values of  $A^{FB}$ and  
${\bar A}^{FB}$ are shown as functions of $\lambda$. 
 The difference $A^{FB} -{\bar A}^{FB}$ is strongly dependent 
on $\lambda$.  
For unpolarized beams one needs around 5\% accuracy of measurements, 
while for $\lambda =\pm 0,5$, an accuracy of 10\% would be enough 
to disentangle $A^{FB}$ from $\bar A^{FB}$. Note that the value 
of this asymmetry is rather large -- even for 
unpolarized $t$-quarks it is about 20\%. 
 
On Table 1 we draw attention that for unpolarized beams  
the uncertainties in  $m_t$ might not allow to make clear 
predictions about the $t$-polarization -- the $m_t$-dependence  
can mimick the $\xi$-dependence of 
 $A^{FB}$. For example, at $\lambda =0$ we have  
  $\bar A^{FB}(m_t =165 GeV) = A^{FB}(m_t =185 GeV)$. 
The beam polarization would allow 
 to distinguish the decay of polarized and 
depolarized $t$-quarks independently on the uncertainties in $m_t$: 
at $\lambda = -0.5$, $A^{FB}-\bar A^{FB} \geq 10\%$ for any value of 
$m_t$. 
 
Fig.3 shows the dependence of ${\cal A}^{FB}$ and 
 ${\bar{\cal A}}^{FB}$  
on $\lambda$ for $m_t=175 GeV$. It is seen that  
the difference ${\cal A}^{FB}-{\bar{\cal A}}^{FB}$ does 
 not exceed 5\%. As shown in Table 1  the uncertainties  
in $m_t$ can smear this difference and higher precision for $m_t$ 
would be necessary to use this asymmetry for to determination of  
the $t$-polarization.  
 
The polarization asymmetry ${\cal B}^{FB}$ appears to be  
the most sensitive one to $t$-quark polarization - the difference  
 ${\cal B}^{FB}-{\bar{\cal B}}^{FB}$ is of  about 40\% and it is not 
reduced by the uncertainties in $m_t$ (see  Table 1).  
The magnitude of the asymmetry is also quite large:   
for $m_t=175$ GeV, ${\cal 
B}^{FB} \approx -0,74$.  
%%%%%%%%%%%%%%%%%%%%%%%%%%%%%%%%%%%%%%%%%%%%%%%%%%%%%%%%%%%%%%%%%%% 
 
\vspace{1.00cm} 
{\bf 7.} We are thankful to Serguey Petcov for the helpful remarks 
and discussion. This work was supported by the 
Bulgarian National Science Foundation, Grant 510 . 
 
\newpage 
 
\begin{table} 
\begin{tabular}{|c||c|c||c|c||c|c|} 
\hline \hline 
$m_t$ & $A^{FB}(0)$ & $\bar A^{FB}(0)$ & $A^{FB}(0.5)$ 
& $\bar A^{FB}(0.5)$ & $A^{FB}(-0.5)$ &  
$\bar A^{FB}(-0.5)$ \\ \hline\hline 
165 & 0.29 & 0.25 & 0.22 & 0.29 & 0.35 & 0.23 \\ \hline 
175 & 0.28 & 0.22 & 0.16 & 0.24 & 0.34 & 0.21 \\ \hline 
185 & 0.25 & 0.19 & 0.12 & 0.20 & 0.34 & 0.18 \\ 
\hline \hline 
$m_t$ & ${\cal A}^{FB}_{LR}(0.5)$ & $\bar {\cal A}^{FB}_{LR}(0.5)$ &  
${\cal A}^{FB}_{LR}(-0.5)$ & $\bar {\cal A}^{FB}_{LR}(-0.5)$ &  
${\cal B}^{FB}$ & $\bar {\cal B}^{FB}$ \\ \hline\hline 
165 & 0.00 & -0.05 & 0.52 & 0.52 & -0.63 & -0.28 \\ \hline 
175 & -0.03 & -0.08 & 0.50 & 0.50 & -0.74 & -0.28 \\ \hline 
185 & -0.06 & -0.12 & 0.47 & 0.47 & -0.86 & -0.28 \\ \hline\hline 
\end{tabular} 
 
\caption{The dependence of the angular asymmetries on $m_t$, 
 in brackets is the beam polarization 
$\lambda =0; \pm 0,5$, ${\cal A}^{FB}(0) = A^{FB}(0)$,   
${\cal B}^{FB}(0)=0$ } 
\end{table}

\vspace{1.00cm} {\bf Figure Captions}  \begin{enumerate} 
\item[{\bf Fig.~1:}] The energy asymmetry $R(E_{max}-E_{min})$ as 
a function of the beam polarization $\lambda$, $m_t$ = 175 GeV. 
\item[{\bf Fig.~2:}] The asymmetry $A^{FB}$ for polarized (full 
curve) and depolarized (dashed curve) $t$-quarks as a function of 
$\lambda$, $m_t$=175 GeV. 
\item[{\bf Fig.~3:}] The asymmetry ${\cal A}^{FB}$ for polarized (full 
curve) and depolarized (dashed curve) $t$-quarks as a function of 
$\lambda$, $m_t$=175 GeV. 
\end{enumerate}    

\newpage
\begin{center}
\setlength{\unitlength}{1mm}
\begin{picture}(150,150)
\put(  0,-15){\mbox{\epsfig{file=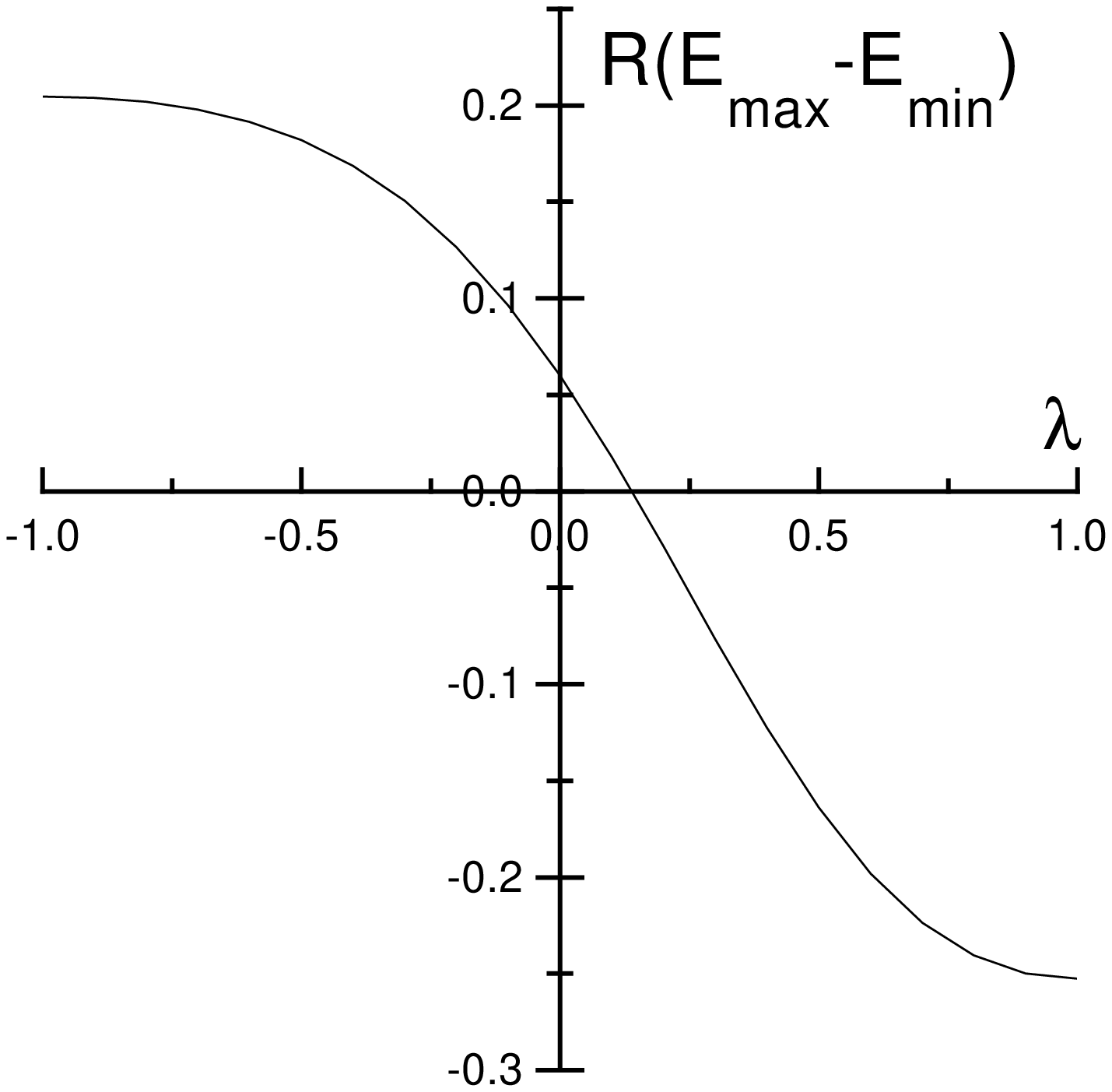,width=15cm}}}
\put( 75, 0){\makebox(0,0)[b ]{{\large\bf Fig.~1}}}
\end{picture}\\
\setlength{\unitlength}{1pt}
\end{center}
\newpage
\begin{center}
\setlength{\unitlength}{1mm}
\begin{picture}(150,150)
\put(  0,-15){\mbox{\epsfig{file=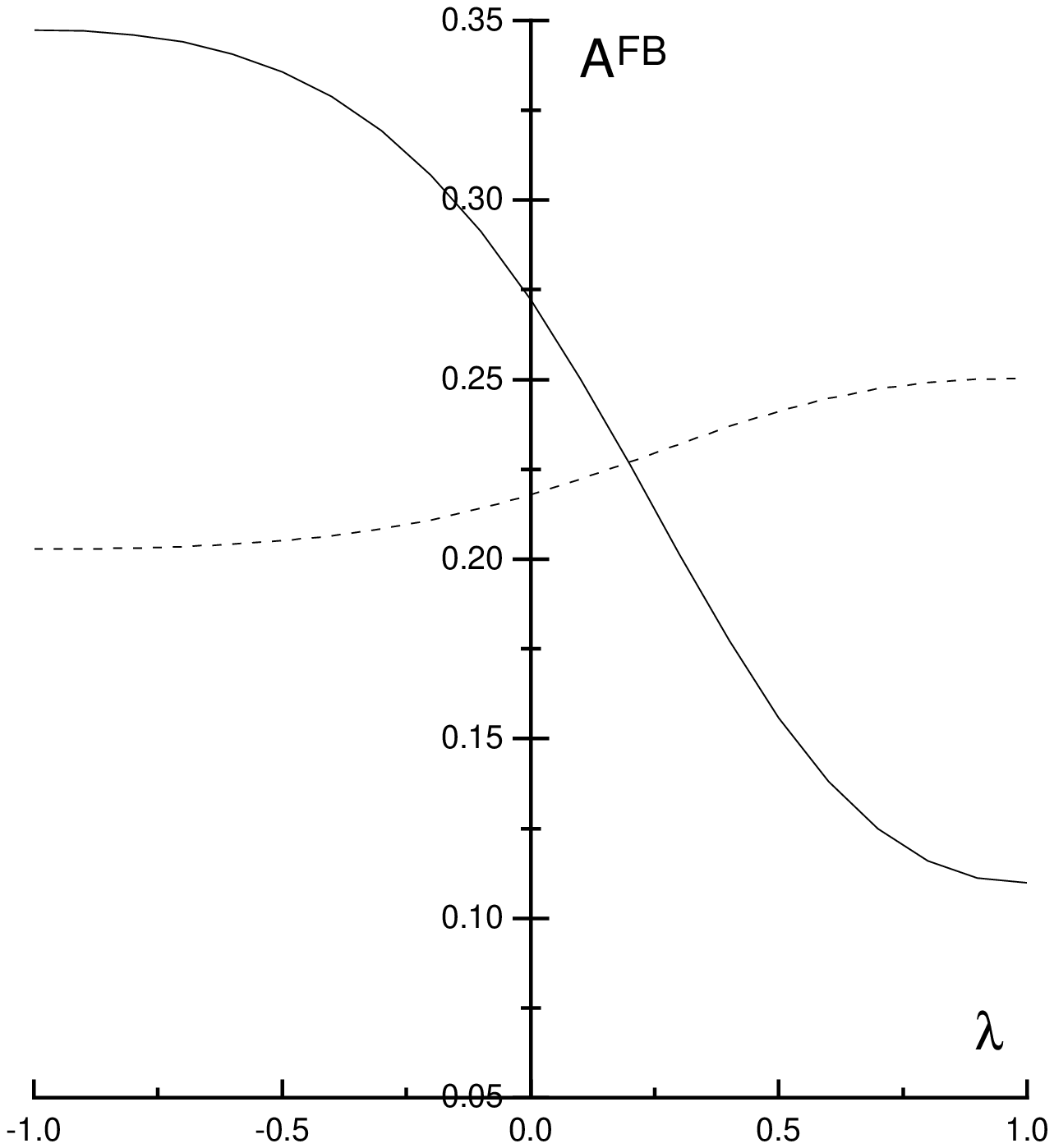,width=15cm}}}
\put( 75, 0){\makebox(0,0)[b ]{{\large\bf Fig.~2}}}
\end{picture}\\
\setlength{\unitlength}{1pt}
\end{center}
\newpage
\begin{center}
\setlength{\unitlength}{1mm}
\begin{picture}(150,150)
\put(  0,-15){\mbox{\epsfig{file=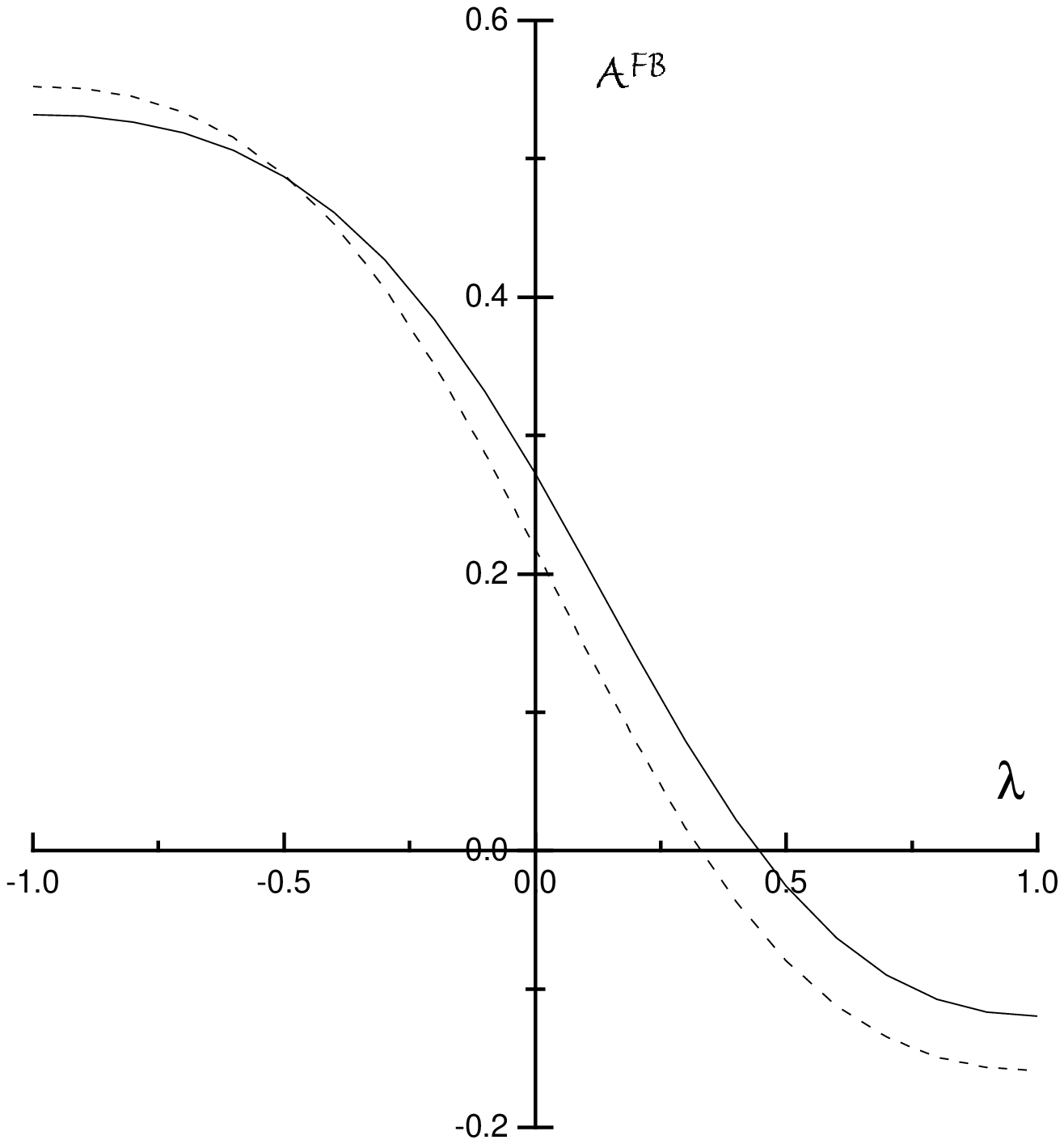,width=15cm}}}
\put( 75, 0){\makebox(0,0)[b ]{{\large\bf Fig.~3}}}
\end{picture}\\
\setlength{\unitlength}{1pt}
\end{center}

\end{document}